\documentclass[11pt]{article}
\usepackage[a4paper,margin=1in]{geometry}
\usepackage{hyperref}
\usepackage{graphicx}
\providecommand{\Description}[1]{}
\usepackage{amsmath,amssymb}
\usepackage{booktabs}

\usepackage[numbers,sort&compress]{natbib}
\title{Beyond Words: Infusing Conversational Agents with Human-like Typing Behaviors}
\author{
  Jijie Zhou\\
  Independent Researcher\\
  Boston, MA, USA\\
  \texttt{seazon96zhou@gmail.com}
  \and
  Yuhan Hu\\
  Independent Researcher\\
  San Jose, CA, USA\\
  \texttt{yuhan3663@gmail.com}
}
\date{}

\begin{document}
\maketitle

\begin{center}\small
\textbf{Author's version.} The Version of Record appears in \emph{Proceedings of the ACM Conversational User Interfaces 2024 (CUI '24)}, July 8--10, 2024, Luxembourg, Luxembourg.\\
DOI: \href{https://doi.org/10.1145/3640794.3665560}{10.1145/3640794.3665560}.
\end{center}

\begin{abstract}
Recently, large language models have facilitated the emergence of highly intelligent conversational AI capable of engaging in human-like dialogues.
However, a notable distinction lies in the fact that these AI models predominantly generate responses rapidly, often producing extensive content without emulating the thoughtful process characteristic of human cognition and typing.
This paper presents a design aimed at simulating human-like typing behaviors, including patterns such as hesitation and self-editing, as well as a preliminary user experiment to understand whether and to what extent the agent with human-like typing behaviors could potentially affect conversational engagement and its trustworthiness.
We've constructed an interactive platform featuring user-adjustable parameters, allowing users to personalize the AI's communication style and thus cultivate a more enriching and immersive conversational experience. 
Our user experiment, involving interactions with three types of agents—a baseline agent, one simulating hesitation, and another integrating both hesitation and self-editing behaviors—reveals a preference for the agent that incorporates both behaviors, suggesting an improvement in perceived naturalness and trustworthiness.
Through the insights from our design process and both quantitative and qualitative feedback from user experiments, this paper contributes to the multimodal interaction design and user experience for conversational AI, advocating for a more human-like, engaging, and trustworthy communication paradigm.

\end{abstract}


\section{Introduction}

\begin{figure}[t]
    \centering
    \includegraphics[width=0.45\textwidth]{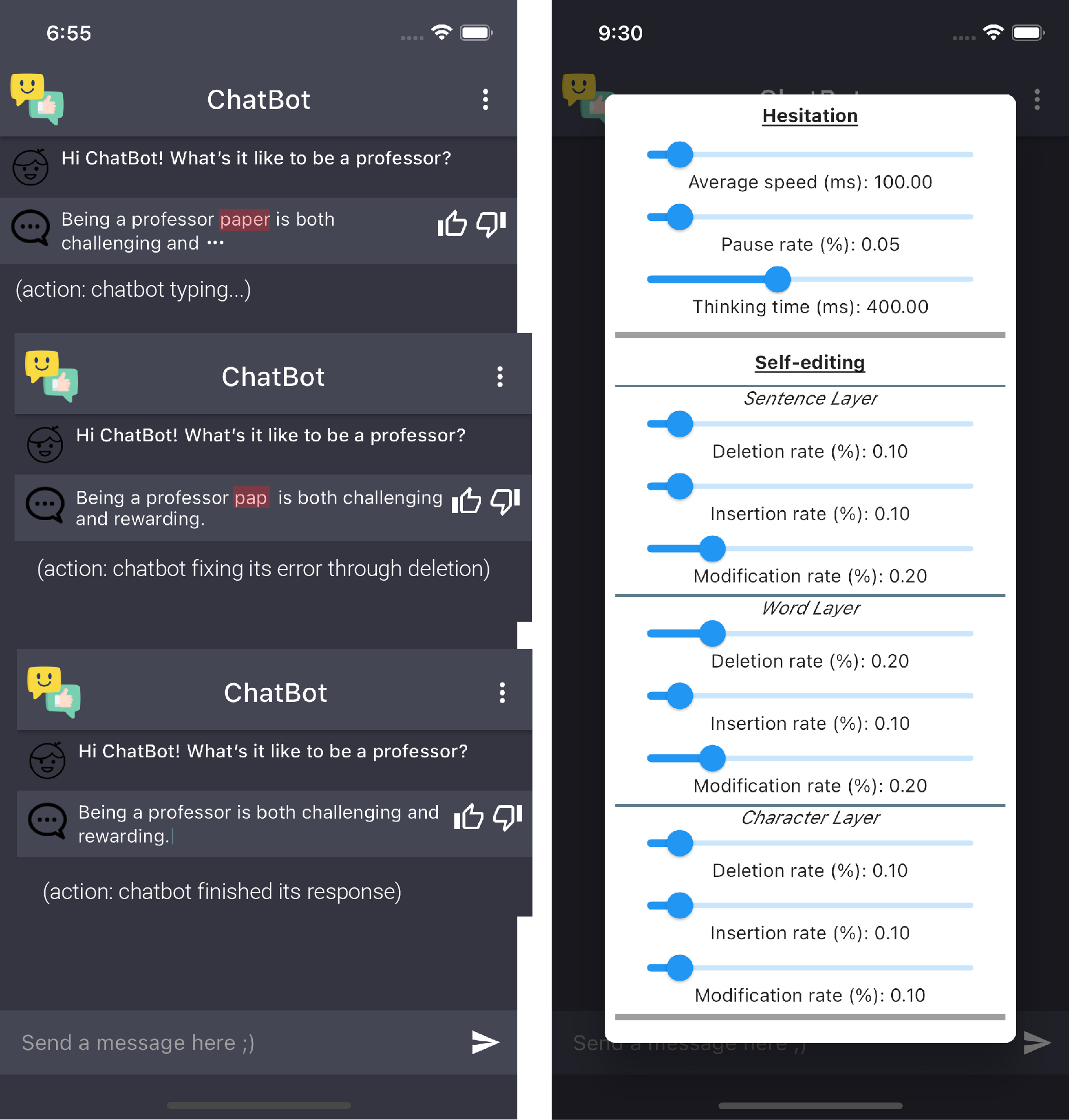}
    \caption{User Interface Screenshot: Demonstrating the chatbot's typing behavior with self-editing capabilities (left), and adjusting parameters in ChatGPT's response (right). In our user experiment, the parameters are pre-defined for each agent.}
    \label{fig:enter-label}
    \Description{The left screenshot illustrates the chatbot's typing behavior with self-editing capabilities where the chatbot corrects a typographical error. The right screenshot shows an interface for adjusting various parameters in ChatGPT's response, including hesitation, self-editing at three different layers (sentence, word, character).}
\end{figure}

ChatGPT has recently emerged as a sophisticated conversational AI, leveraging large language models \cite{chatgpt}. Numerous studies have indicated that it can be challenging for humans to discern whether they are interacting with an AI or another human~\cite{cai2023does, dwivedi2023so}. 
However, there is one telltale aspect that potentially distinguishes ChatGPT from human interlocutors: its typing behaviors. Unlike humans, who pause to think, make typos, or correct their errors, ChatGPT produces responses with remarkable speed and precision, exhibiting virtually no signs of hesitation or errors. This element of its communication style could be a significant identifier for those trying to differentiate between AI and human correspondences.

Numerous studies suggest that a person's typing behavior can be a rich source of information about their mental state, encompassing aspects such as emotions, cognitive processes, confidence, and interests~\cite{lee2014influence, shikder2017keystroke}. Drawing from the principles of multimodal interaction design, this paper aims to imbue ChatGPT with human-like typing behaviors, thereby enhancing the user experience by simulating more natural, human-like interactions. By introducing an interactive platform that uses ChatGPT as its core technology and simulates behaviors typically associated with human typing, such as hesitation and self-editing, we add an additional layer of human-like complexity to the AI's communication style.

By incorporating a variety of user-adjustable parameters into the output of the chatbot's typing behaviors, individuals can readily modify the communication style to display various characteristics, such as the level of confidence or hesitation in responses. This method enables an enhancement of the interaction by incorporating elements akin to non-verbal cues, thereby enriching the overall communication experience.

This paper conducts a within-subject study to examine users' perceptions of ChatGPT-based conversational agents, focusing on three variations that utilize the same content generation model but differ in typing behaviors. The three agents under comparison are: a baseline agent (blue), another with integrated hesitation behavior (green), and a third that combines both hesitation and self-editing behaviors (red). Analysis of users' quantitative and qualitative feedback reveals that the agent exhibiting both hesitation and self-editing behaviors (red) is the most preferred among participants. Users describe this agent as more “natural”, “human-like”, and “thoughtful” in generating responses, with its self-editing capability particularly appreciated as an indicator of intelligence and error-correction ability. Conversely, the baseline agent (blue) is often perceived as “unreal” due to its rapid typing speed, leading many participants to this conclusion. Meanwhile, the agent with hesitation behavior alone (green) is the least favored, with participants deeming it “dumb” and “robotic”.

We aim for this research to serve as a catalyst for further design innovations and the formulation of design recommendations within the domain of ChatGPT and AI agent development. Our focus extends beyond the content-generation layer to also consider the delivery mode of the content to humans across diverse social contexts. Our objective is to create more engaging, contextually aware, and ultimately, more human-like conversational AI experiences.

\section{Related Work}

In this section, we survey existing literature pertaining to user perception of chat agents with varied typing behaviors and character embodiment. Furthermore, we provide an overview of the development and evolution of applications based on Large Language Models, establishing the foundation for our research.

\subsection{Typing/Character Perception}
% error
Research findings indicating that people interact with intelligent machines in ways similar to their interactions with humans have inspired researchers to explore interactions between humans and computers or robots, with the goal of improving human trust, affection, and overall efficiency in robot usage~\cite{Nass2000MachinesAM}. Studies involving embodied chat robots reveal that perceived attention and intelligence are highly correlated with animacy~\cite{Bartneck2009DoesTD}. Moving towards disembodied chat agents, human-like traits have proved to impute more positive characteristics into chatbot~\cite{Appel2012DoesHM}. Perception of animacy in the interactions with disembodied chat agents can be obtained from socially oriented conversations such as including greetings and emotional reactions to users statements~\cite{Laban2021PerceptionsOA}. Appearances like using human-like avatars, conversational tone, emotional message cues are also major anthropomorphic cues that have been shown to improve the disembodied chatbot's perceived trustworthiness, intelligence, and usefulness~\cite{Donkelaar2018HowHS, Cai2022AnthropomorphismAO}. Existing research has primarily focused on textual and lexical properties in chatbot dialogues when exploring the concept of animacy. However, there is a notable gap in the literature concerning the impact of visual cues in displaying dialogues, which can significantly influence the perceived human-likeness of disembodied chatbots. Although there are graphical and textual visual cues to indicate that the chatbot is preparing an answer, typing indicators and perceived social presence of chatbots still highly depend on user’s prior experience with chatbots~\cite{Gnewuch2018TheCI}. These cues provide indications before showing the message, rather than revealing the entire flow of generating every single part of a sentence. As an example, the presentation of typing dynamics, implicitly conveying the thought process behind text generation, may significantly contribute to enhancing the animacy of chatbots. Research has shown that typing behaviors can reveal substantial information about an individual's mental state and are perceptible to other humans. For instance, Vizer et al.~\cite{vizer2009detecting} demonstrated that typing behavior could indicate stress levels and cognitive load, while Ong et al. \cite{ong2018qwerty} found a correlation between typing speed and topic familiarity in search queries. Cohen et al. \cite{cohen1996behavioral} observed the momentum of typing behavior in competitive tasks, such as resistance to making changes. Inspired by these findings, this paper investigates two specific typing characteristics: self-correction and temporal delays in typing.

\subsubsection{Self-correction}

\begin{sloppypar}
While certain human-like characteristics have been established as effective strategies to enhance chatbot performance, there remain unverified traits, such as chatbots making mistakes during interactions with users, whose impact on effectiveness is yet to be firmly established. Previous research in chatbot interactions has addressed aspects such as typing errors, user satisfaction, and human-likeness perception. They provide insights into the influence of typing errors on perceived humanness and service quality~\cite{Bhrke2021IsMM}. Most of these studies have revealed that typing errors lead to lower perceived humanness and social presence, while certain human-like cues may simultaneously increase satisfaction and frustration~\cite{Brendel2023ThePR}. Studies tend to indicate that mistakes, such as typos and capitalized words, do not improve user perception regarding the long-term utilization of robots like conversational agents~\cite{Westerman2019IBI, Brandtzg2018ChatbotsCU}. Some also suggest that a generally human-like communication style negatively affects the perceived efficiency of chat agents~\cite{deSSiqueira2023WhenDW}. However, it's important to note that various experimental conditions may influence participants' ability to discern whether they are interacting with a human or a robot. Conditions such as pre-generated sets of questions~\cite{Westerman2019IBI} or predefined service topics~\cite{Bhrke2021DoYF} can potentially bias users and shape their expectations, leading to assumptions about the responses they receive and the true identity of the agent they are chatting with. 
\end{sloppypar}

While correction behavior in chatbots has been studied extensively, with a focus on stages where they perceive user queries~\cite{Arifi2019PotentialsOC, Ergen2023ContextAS}, existing research has predominantly centered on content-related aspects of response generation. However, an area within human-like traits that has received limited attention is self-correction. The studies we've introduced have primarily explored how users react to chatbots making mistakes, but there is a noticeable scarcity of experiments involving chatbots proactively self-correcting their errors during the typing process and transparently displaying this correction to users.

\subsubsection{Temporal Delay}
The perception of a robot's human-likeness can be influenced by the temporal dimension of its responses. Previous studies have demonstrated that chatbots sending static~\cite{Appel2012DoesHM} or dynamically delayed responses, calculated based on the complexity of the user's message and the generated text, were perceived as more human-like and exhibited higher social presence compared to chatbots sending near-instant responses~\cite{Gnewuch2018FasterIN}. However, these results were obtained when participants were informed that they were interacting with a non-human robot agent~\cite{Gnewuch2018FasterIN, Gnewuch2022OpposingEO}. There is a need for further research and theoretical support for scenarios where user awareness of human or robot interaction is intentionally ambiguous, with the aim of enhancing the efficiency of chatbot usage.

Other research has indicated that response time variability positively affects users' inclination to follow chatbot advice~\cite{Milana2023ChatbotsAA}. However, these studies typically focus on delays in preparing entire sentences, rather than examining delays before individual words or characters. Additionally, there is currently no research that combines the temporal dimension (delay) with the textual dimension of robot animacy, such as the occurrence of mistakes or typos in generated text.

% hesitaiton

\subsection{ChatGPT/LLM-based and Conversational AI applications}

The first generation of chatbots, exemplified by ELIZA~\cite{Weizenbaum1966ELIZAaCP}, aimed to simulate the responses of a psychotherapist during therapy sessions, employing a pattern matching and substitution methodology. During this era, when computer-mediated conversations were still a novelty for users~\cite{Adamopoulou2020AnOO}, such chatbots successfully perplexed and engaged people. Subsequently, researchers began employing natural language models to construct chat agents capable of delivering human-like services. A.L.I.C.E. utilized AIML (Artificial Intelligence Markup Language) and rule-based scripting to generate responses~\cite{Marietto2013ArtificialIM}, while Jabberwacky adopted a learning-based approach, adapting its responses based on user interactions.

Among modern personal assistants, Apple Siri~\cite{siri} employs Natural Language Understanding (NLU), while IBM Watson~\cite{watson} relies on Natural Language Processing (NLP) to efficiently carry out practical tasks. Over time, chatbot responses have evolved, primarily involving the generation of responses using machine learning techniques, often from scratch, employing generative models~\cite{Nirala2022ASO}. AI-powered chatbots have now found widespread application in diverse fields, including healthcare~\cite{Nadarzynski2019AcceptabilityOA}, public service~\cite{Nirala2022ASO}, and education~\cite{Haristiani2019ArtificialI}.

As the latest development in this group, ChatGPT, an online large language model (LLM), was introduced in November 2022. This generative model is unsupervised, and pre-trained, and due to the vast amount of unlabeled data used in training, it has earned the distinction of being one of the largest language models in the world~\cite{Floridi2020GPT3IN}. Through a comprehensive analysis of GPT-3.5~\cite{gpt35} and its predecessors, it becomes evident that GPT-3.5 outperforms existing fine-tuned models, although further enhancements to its robustness may still be necessary~\cite{Ye2023ACC, Chen2023HowRI}. Notably, GPT-3.5 excels in terms of processing speed, rendering it exceptionally well-suited for real-time applications, such as chatbots~\cite{Rahaman2023FromCT}. Additionally, its proven helpfulness in responses positions ChatGPT as a strong contender for serving as a personal assistant~\cite{Guo2023HowCI}. It should be noted, however, that careful prompt development is essential to achieve the best results with ChatGPT, as it has limitations~\cite{Lee2023BenefitsLA}. Nevertheless, we believe that these limitations can be leveraged for applications where specific conditions or hidden constraints are necessary for obtaining responses from ChatGPT. Its sensitivity to prompts enables us to finely shape the responses we desire, including constraining the language of response to meet specific user needs.

\begin{sloppypar}
Building on these advancements in multimodal conversational AI, the field continues to evolve, notably with the introduction of ChatGPT. Recent research has highlighted the importance of integrating multimodal interactions, such as video, audio, and text, to accurately capture user intentions and improve dialogue strategies, thereby enhancing user intention recognition and satisfaction~\cite{Yu2021IncorporatingMS}. Further development of datasets like SIMMC 2.0~\cite{Kottur2021SIMMC2A} stresses the importance of embedding users' multimodal context within task-oriented dialog systems for generating more effective responses. Conversely, on the output side, advancements emphasize the AI's capability to provide multimodal information, with AiCommentator and SIVA exemplifying the use of visual content and expressive responses to offer richer and more engaging user interactions~\cite{Andrews2024AiCommentatorAM, Aneja2021UnderstandingCA}. This shift towards AI's multimodal output capabilities, especially our study's focus on visually depicting typing behavior, aims to enhance the interactivity and naturalness of human-AI communication, exploring a novel aspect of multimodal conversational AI.
\end{sloppypar}

\section{Design}

% In this section, we introduce the foundational elements of typing behaviors, along with potential parameters for variation within the behavior-generation platform. 
%These components serve as the building blocks in our endeavor to simulate more human-like conversational experiences in AI.

\subsection{Design Space}

We propose a design framework that encapsulates human-like typing behaviors, specifically modeling hesitation and self-editing patterns typically exhibited in human typing. 

The hesitation facet of the design space incorporates adjustments to the overall typing speed to simulate contemplation, and the intermittent insertion of pauses to mirror human thought processes. The parameters encompass adjustments to both the average and variance of typing speed, the frequency of hesitations (represented as pause rate), and the duration of each pause, measured in seconds.

When it comes to self-editing, we outline three behavioral patterns: deleting, inserting, and modifying previously typed text. Each self-editing behavior can be quantified with a particular rate, representing the frequency with which that behavior might occur, and this is measured in terms of probability.

\begin{figure}[t]
    \centering
    \includegraphics[width=0.5\textwidth]{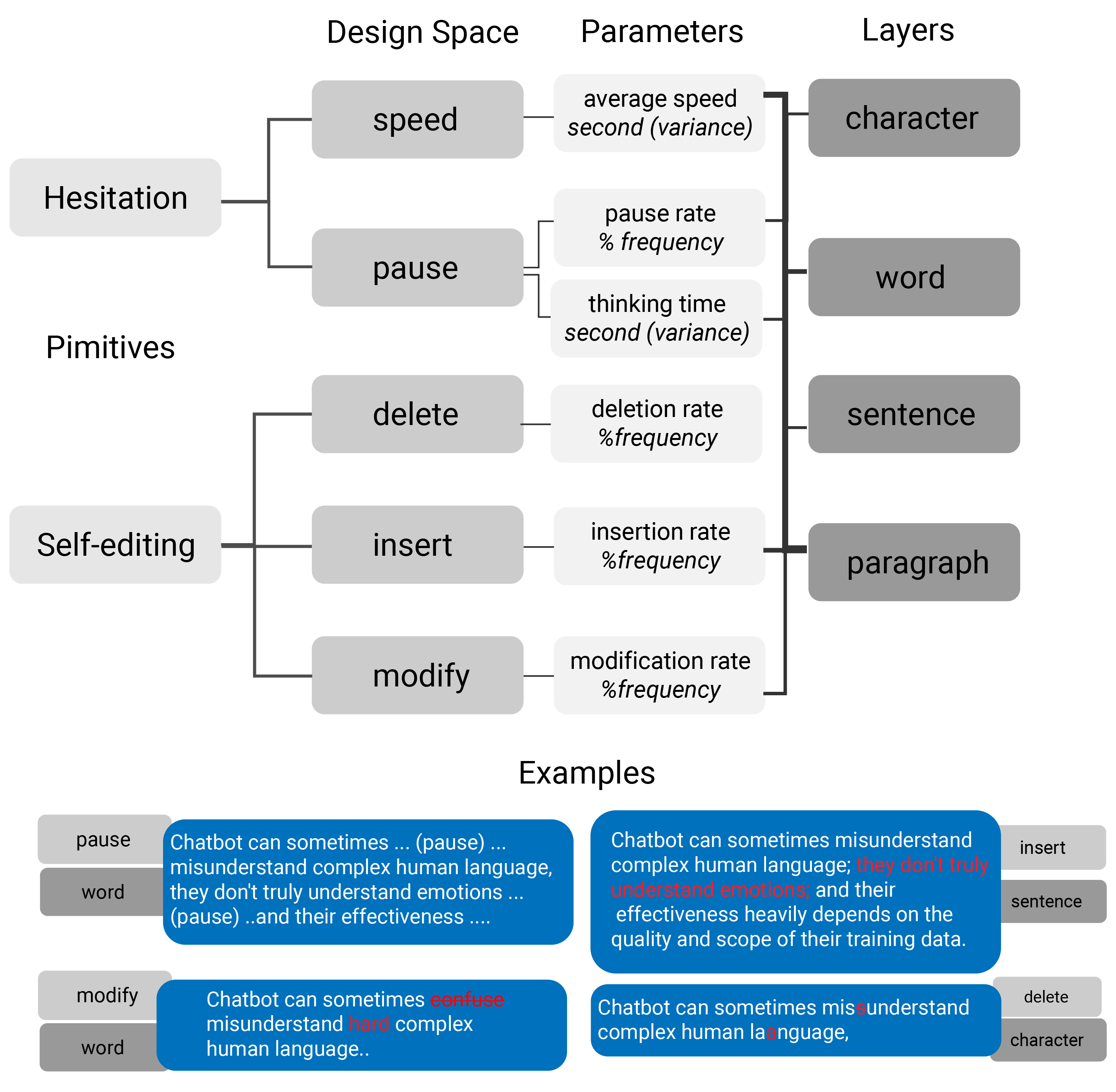}
    \caption{Design space and parameters of human-like typing behaviors with examples.}
    \label{fig:design_space}
    \Description{The figure illustrates the design space and parameters of human-like typing behaviors, divided into two categories: hesitation, and self-editing. The parameters include speed, pause, delete, insert, and modify, with associated metrics such as average speed, pause rate, thinking time, deletion rate, insertion rate, and modification rate. Each parameter is mapped to different layers, such as character, word, sentence, and paragraph. Examples are provided to demonstrate how these behaviors are applied in practice.}
\end{figure}

These elements of editing and hesitation can be manifested at four distinct layers: the character level (e.g., rectifying a typo), the word level (e.g., fine-tuning terminology), the sentence level (e.g., adding clarifying sentences), and the paragraph level (e.g., exhibiting hesitation prior to the completion of a paragraph). Figure~\ref{fig:design_space} depicts the framework of the proposed design space, illustrating example behaviors that can be generated within this structure.

\subsection{Platform Implementation}

We've integrated our platform using Flutter~\cite{flutter}, an open-source app development framework. When a user poses a question, we utilize the OpenAI API to submit the query, simultaneously applying certain constraints, such as the anticipated length of the response. Upon receiving the response from the API, we tailor its presentation to align with designated typing patterns, which may include introducing hesitation delays or selecting phrases for self-editing.

Regarding self-editing behavior, we utilize a rate parameter set via the user interface to randomly select a sentence, word, or character for deletion, modification, or insertion. Each element is edited only once. For modifications, we leverage the WordsAPI~\cite{wordsAPI} to identify a synonym for the word in question. For deletions, we select redundant words from our own vocabulary library, which contains common words used in daily chats.
During each display cycle, our framework processes responses by preparing initial text, a sequence of modification actions, and the final text for user rendering. We also implemented widgets in the rendering stage to allow users to observe the process of typing. As the chatbot types, a flashing cursor tracks the text insertion position, illustrating the real-time typing sequence.

The platform features several features:

\textbf{Flexibility}: Our platform offers various adjustable parameters for each typing behavior, encompassing the behavior's frequency, timing parameters, and the level at which it occurs, whether character, word, or sentence. This level of customization enables us to emulate a diverse range of typing behavior outcomes.

\textbf{Immediacy}: Users or platform operators can modify parameters at any point during a conversation. Adjustments take effect instantaneously, ensuring a dynamic and responsive user experience.

\textbf{Randomness}: The self-editing behavior is randomly generated, implying that any action could occur on any part of the sentence at any time. For instance, a sentence-level deletion could transpire as the chatbot is currently typing, or it could occur retrospectively after the entire paragraph has been written.

\textbf{Controllability}: When the system decides what and when to delete, it allows for certain sentence components to remain unchanged. For example, the start of the sentence is immune from certain word modifications, such as deletion, mimicking the general human typing behavior where initial sentence elements are less likely to be altered.

\textbf{Visibility}: We allow users to choose the mode to display the typing process of each message from the agent. This option gives people chances to witness the agent's whole typing process. 

\textbf{Cross-Platform}: Our platform can be accessed from either web or mobile platforms including iOS and Android.

\textbf{Open-Source}: Our platform is open-source and accessible on GitHub: \url{https://github.com/jigglypuff96/human-like-typing-bot}

\subsection{Design Parameters}
Below we outline the implementation details and the parameters that can be adjusted for each typing behavior.

\subsubsection{Temporal Parameters}

We can simulate human-like typing patterns to emulate the behavior of hesitation by moderating the typing speed, incorporating non-uniform short pauses, and introducing extended breaks that mimic periods of contemplation. As illustrated in the accompanying video, designers or users can adjust these parameters to create responses tailored to their reading preferences. This approach can give the impression that the chatbot is engaging in a thoughtful process, thereby enhancing its perceived human-like quality. 
%However, it's worth noting that such behavior could also be interpreted as a sign of uncertainty or lack of confidence in the chatbot's abilities.

We established a set of key parameters to control the chatbot’s typing behavior effectively. These parameters were instrumental in replicating the nuanced aspects of human typing and interaction.

\textit{Single Character Typing Pace} We introduced the concept of "characterTypingPace," representing the time it takes for the chatbot to type each character of a word. This parameter allowed us to modulate the speed at which the bot typed out the text, creating a dynamic typing rhythm that closely mimicked human variability, including a deliberate and thoughtful typing style when set to a slower character typing pace.

\textit{Space Lag Pace} "spaceLagPace" is a parameter designed to govern the duration of the pause before the chatbot began typing a word, simulating moments of contemplation or uncertainty that are characteristic of human conversation. A longer space lag pace can indicate more profound contemplation.

\textit{Single Character Deletion Pace} Recognizing the importance of addressing errors or revisions in a conversation, we introduced "characterDeletionPace" as another parameter. This parameter controlled the speed at which the chatbot deletes or backspaces characters, simulating the thoughtful correction process humans undertake when they make mistakes or wish to modify their input. A slower deletion pace can imply more careful editing.

\textit{Cursor Moving Speed} Lastly, we incorporated "cursor moving speed" as an additional parameter. This aspect of the model allowed us to replicate the way humans navigate their text when they realize errors or inappropriate content within their messages. Faster cursor moving speed can indicate swift revisions.

We implemented a normal distribution process to model the actual waiting times. Within this framework, the previously defined parameters were subjected to normal distributions. This statistical approach allowed us to introduce randomness and variability into the hesitation patterns, mirroring the unpredictability of human behavior during conversation.

\subsubsection{Self-editing Parameters}

We've introduced several self-editing behaviors, including deletion, insertion, and modification. By allowing the chatbot to self-correct, we can emulate more authentic human-like typing patterns. After all, humans often make mistakes, such as typos and grammatical errors, and tend to self-correct during the writing process, revisiting and revising previous phrases. We hope that, by incorporating behaviors with an error rate and self-editing, ChatGPT can be perceived as less robotic. Additionally, this feature expresses uncertainty in arguments through repeated modifications of terms.

We have established a comprehensive framework that encompasses four distinct levels of parameterization. These parameters serve as the building blocks for emulating self-editing behaviors.

\textit{Paragraph Level} At this level, we offer the flexibility to set the proportion of sentences that will undergo deletion, insertion, or modification. This level of control allows for a macroscopic adjustment of self-editing behavior within the chatbot's responses.

\textit{Sentence Level} Moving to the second level, the sentence level, we introduce the ability to define the proportion of words to be deleted, inserted, or modified within a sentence. It's important to note that all actions at the sentence level are derived exclusively from the paragraph-level modifications, ensuring coherence and consistency within each paragraph.

\textit{Word Level} The third level, the word-level, focuses on actions that occur within the selected sentence chosen for modification. This granular control allows for precise adjustments to the chatbot's self-editing tendencies at the word level.

\textit{Character Level} Lastly, at the character level, we address typographical errors. This level of detail permits fine-tuning of the chatbot's ability to rectify minor errors, further enhancing its human-like text generation.

Given that our platform is tailored for chatbot interactions, typically composed of shorter text snippets, the emphasis lies in configuring the proportions of actions at the latter three levels, as multiple paragraphs are less common in this context.

Importantly, the total proportions of actions at each level do not exceed 1, ensuring that the overall self-editing behavior remains balanced and coherent. Additionally, to maintain consistency and avoid redundancy, the same item will not undergo more than one action. For example, if a word is selected for insertion, it will not be later modified, preserving the natural flow of conversation and self-editing akin to human interactions. This multi-tiered approach to self-editing parameters contributes to the chatbot's ability to exhibit nuanced and convincing self-correction behaviors, ultimately reducing its perceived robotic nature and infusing it with a desirable degree of uncertainty in arguments through repeated term modifications.

\section{Experiment}
In this section, we present a pilot user study to evaluate perceptions of typing behaviors in AI chat companions. This preliminary investigation seeks to initiate deeper research into conversational agents' hesitation and self-editing behaviors. By acknowledging this study's exploratory nature upfront, we highlight our commitment to advancing conversational AI design and underline the need for future studies with larger participant groups to solidify our findings.

\subsection{Research Questions and Hypothesis}
Our study is driven by two major research questions.

\textit{RQ1: Whether and to what extent will the incorporation of hesitation behaviors into the chat agent lead to perception change of naturalness and likability?}
Our hypothesis posits that the introduction of hesitation behaviors will enhance the agent's human-likeness (\textit{H1}), and elevate the agent's likability among participants (\textit{H2}).

\textit{RQ2: Whether and to what extent will the integration of self-editing behaviors into the chat agent affect its perceived naturalness and likability? }
We hypothesize that implementing suitable self-editing behaviors will make the agent more human-like (\textit{H3}); and will increase the agent’s perceived competence (\textit{H4}).

\subsection{Methodology}
We construct a \textbf{within-subject study} to examine our research questions and hypotheses. This study incorporates three distinct chat agents: a baseline agent labeled \textbf{Blue}, an agent exhibiting hesitation behavior named \textbf{Green}, and an agent that displays both hesitation and self-editing behaviors, referred to as \textbf{Red}. The use of color labels, rather than personal names, aims to focus attention on the behavioral traits of the agents, avoiding any personal biases or assumptions. Each participant engages in a casual conversation with each agent, exchanging up to five messages. Following these interactions, participants are required to complete a questionnaire to share their perceptions of each agent. This questionnaire includes various items, as well as a few open-ended questions to gather qualitative insights. We counterbalance the order in which the agents are presented to mitigate the learning effect.

\subsubsection{Metrics}
\label{sec:metrics}
We employ subjective metrics, collected through participant feedback via questionnaires, and objective metrics used for analyzing the quality and quantity of messages participants exchange with different agents. For each agent involved in the study, participants are asked to assess the \textbf{response time}, \textbf{competence}, \textbf{satisfaction}, \textbf{friendliness}, \textbf{easiness}, \textbf{clarity}, \textbf{naturalness}, \textbf{human-likeness} using a 5-point Likert scale. 
The metrics are designed to assess the quality and efficiency of the user experience, as well as users' perceptions of the agent. These include adaptations of commonly used questionnaires such as the UEQ~\cite{schrepp2015user} and HRIES~\cite{spatola2021perception}, tailored specifically for this chat agent to ensure conciseness and consistency.
Additionally, participants respond to three open-ended questions designed to elicit qualitative input regarding their overall impressions of the agent and suggestions for potential improvements. Upon completion of interactions with all three agents, participants are requested to select the agent they found most enjoyable to converse with and provide reasons for their choice.

Regarding objective metrics, we record the number of words and sentences participants type with each agent. Additionally, we conduct qualitative analyses of the content exchanged with the agents to comprehend the flow of their conversations.

\subsubsection{Task Design}
In an effort to truly grasp the participants' perceptions of the agent within the framework of casual, social chatting, we devised several tasks and used a pre-study to evaluate the task design. Some design candidates included: (1) having participants engage with the three agents, each discussing 3 pre-defined topics; (2) allowing participants to choose the topic of conversation; (3) instructing participants to envision the agent as a newly-met online subject and share their hobbies. Following a series of pilot studies, we decided on the third task design. This approach fosters an experience that leans more towards social awareness than a functional, goal-driven one, providing a context in which participants may be more attuned to the agents' naturalness and human-likeness, rather than focusing solely on the content of the responses. To further minimize any pre-existing biases towards the agents, we refer to each as a "partner" without revealing their identity (i.e., whether they are human or chatbot).

\subsubsection{Interface Design}
Once users enter the chat room to chat with the agent, we will display a floating dialog indicating that they will be connected to a certain agent within a time range (a randomized number selected between 5 to 15 seconds). The delay in connecting to the agent also helps avoid giving people the impression that they are talking to a robot, which is typically available 24/7.

\subsubsection{Participants}
Through snowball sampling via emails and messages, we recruited 20 participants. However, the analysis incorporates data from only 11 of the 20 participants due to incomplete data sets (for example, several participants failed to complete the chat with all the three agents) and one participant failed the attention check, resulting in a total of 33 data points across the three agents. Participant ages, self-reported, ranged from 18 to 34, including 6 males and 5 females. Among them, 5 are native English speakers while the remaining 6 are fluent in English.
% \subsection{Platform}

\subsubsection{Implementations}

When participants engaged with these agents in a chat scenario, instead of sending the participant's chat directly as a query to the agents, we introduced an additional line into the conversation. This additional line, which reads as follows: "(Please provide a reply with no more than x sentences, and less than y words in total. Use English only please.)", was strategically placed to serve a specific purpose. Its function was to guide the response generated by GPT-3.5, using the text-davinci-003 model~\cite{gpt35}, towards a more natural and casual chat-like interaction. By imposing limitations on the number of sentences and words in the response, we aimed to create a conversational dynamic that closely resembled how humans engage in dialogue, characterized by relatively short and concise lines. This approach allowed us to explore the potential of these robot agents to emulate human-like conversational behavior while maintaining control over the response length and language quality. 

\subsection{Procedure}

Prior to the experiment, participants receive an email containing a consent form. This form provides a concise introduction to the experiment, outlines potential risks, and explains the data collection process. Participants are required to sign this form to indicate their consent to participate.
Upon receipt of the signed consent form, each participant is assigned a unique, anonymous User ID. Following this, participants receive a link to an online questionnaire, accessible via PC or mobile device. The initial section of this questionnaire is dedicated to gathering demographic information from the participants.

Following the demographic data collection, participants are introduced to their first task: engaging in a topic exchange with a partner. The partner, selected randomly, is represented by one of three possible colors: Blue, Green, or Red. Participants are instructed to share and discuss hobbies, choosing from categories such as sports, music, activities, food, or movies, all within the confines of five messages.
To initiate the task, participants click a link directing them to a chat platform where the exchange will take place. Notably, participants are kept unaware of the nature of their partner, who may be either a human or an AI. The inclusion of a waiting room feature prior to the connection further obfuscates the identity of the partner, enhancing the believability of the partner as human being. Our design mirrors scenarios where AI's role is undisclosed to encourage natural user interaction, addressing AI skepticism and its impact on engagement and efficiency. Such contexts include online banking, where customers interact with AI for account inquiries without explicit disclosure, and virtual health advisors, providing personalized health guidance. This approach not only enhances the study's ecological validity but also examines AI's potential in roles traditionally filled by humans, reflecting its application in areas like customer service and personal support.

Upon completion of the chat with the first partner, participants are directed back to the questionnaire to respond to several queries evaluating their chat experience. This process is repeated three times in total, with participants interacting with a different, randomly-selected agent each time.
Finally, participants are prompted to answer post-experiment questions, which are designed to elicit direct comparisons between the experiences with the three different agents. Each participant's engagement in the experiment concludes upon completion of this section of the questionnaire.

\section{Result}
A mixed-method approach was employed to collect both subjective and objective data from users, utilizing metrics outlined in Section \ref{sec:metrics}. For the quantitative aspect of our analysis, we compared data derived from questionnaires and quantifiable inputs obtained during users’ conversational exchanges. Concurrently, qualitative analysis was performed to identify themes arising from responses to open-ended questions in the questionnaires.

\subsection{Quantitative Analysis}

\subsubsection{Questionnaire Responses}
\begin{figure*}[t]
    \centering
    \includegraphics[width=0.7\textwidth]{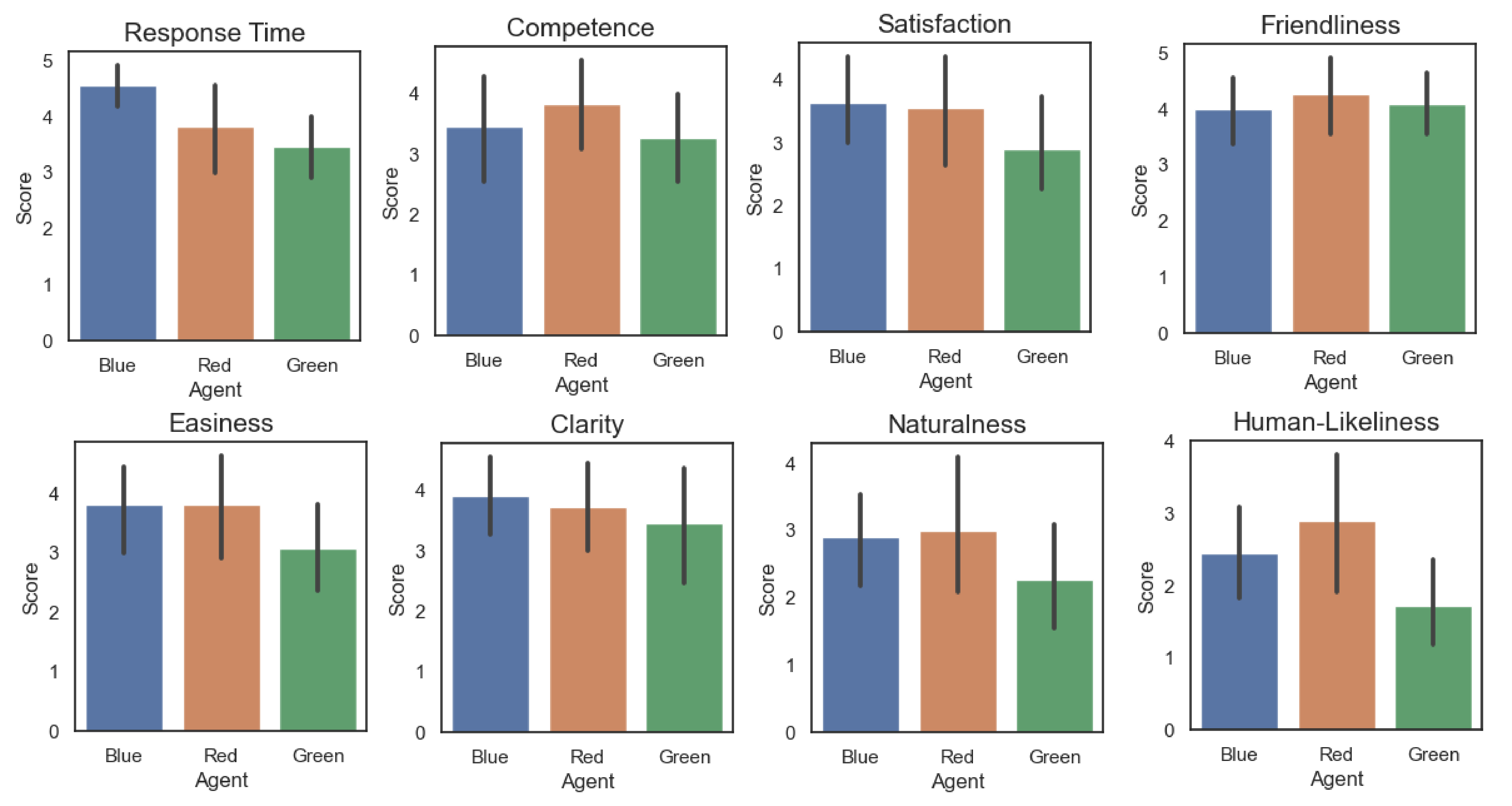}
    \caption{Results of questionnaire responses.}
    \label{fig:survey}
    \Description{The figure shows eight bar graphs representing questionnaire responses on various metrics: Response Time, Competence, Satisfaction, Friendliness, Easiness, Clarity, Naturalness, and Human-Likeness. Each graph compares scores for three different agents labeled Blue, Red, and Green.}
\end{figure*}

Upon engaging with each agent, participants evaluated their experiences using the metrics specified in \ref{sec:metrics}, providing ratings on a 5-point Likert scale. Figure \ref{fig:survey} illustrates the mean score assigned to each metric, with the standard deviation denoting variance.

Concerning \textbf{response time}, the baseline agent (blue $M=4.550, std=0.688$) excels over agents embodying hesitation and self-editing (red $M=3.818, std=1.401$) and just hesitation behaviors (green $M=3.455, std=1.036$). The agent exhibiting only hesitation behaviors (green) generally receives less favorable assessments, notably in \textbf{naturalness} (green: $M=2.272, std=1.272$) and \textbf{human-likeness} (green: $M=1.727, std=1.103$). \textit{H1} is not supported, and \textit{H2} is not supported.

Conversely, the agent incorporating both hesitation and self-editing behavior (red) slightly surpasses the baseline agent (blue) in facets such as \textbf{naturalness} (red: $M=3.0, std=1.789$; blue: $M=2.909, std=1.300$), \textbf{human-likeness} (red: $M=2.909, std=1.701$; blue: $M=2.455, std=1.214$), \textbf{competence} (red: $M=3.818, std=1.401$; blue: $M=3.454, std=1.572$), and \textbf{friendliness} (red: $M=4.273, std=1.272$; blue: $M=4.0, std=1.095$), which supports \textit{H3} and \textit{H4}. However, the observed differences were not deemed statistically significant upon inferential analysis.

\subsubsection{Text entry}

\begin{figure*}[t]
    \centering
    \includegraphics[width=1.0\textwidth]{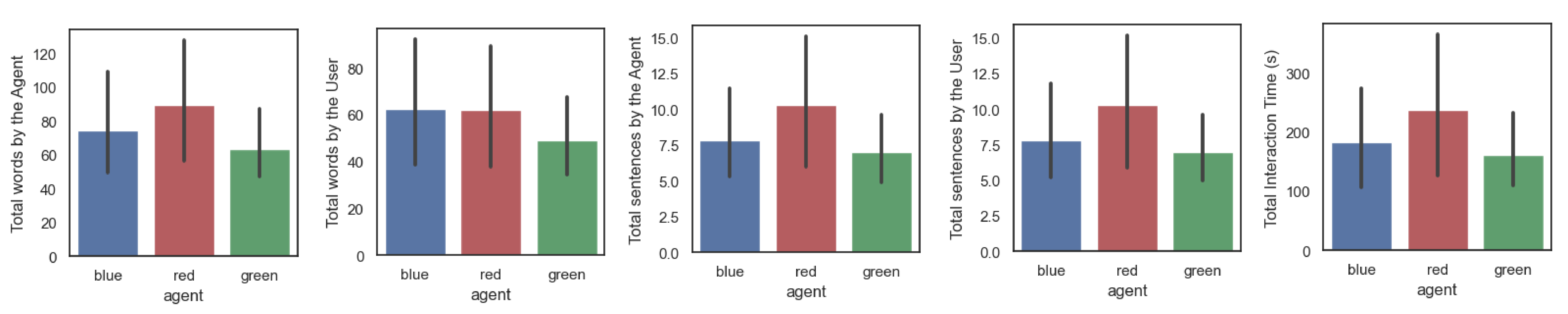}
    \caption{Results of the count of word entry, sentence entry, and interaction duration time by the user and by the agent.}
    \label{fig:compare_word}
    \Description{The figure displays five bar graphs comparing the count of word entries, sentence entries, and interaction duration time between the user and the agent across three agents: Blue, Red, and Green. The metrics include total words by the agent, total words by the user, total sentences by the agent, total sentences by the user, and total interaction time in seconds.}
\end{figure*}

To assess the extent of message exchanges with the agent, we analyze the number of words entered, sentence entries, and overall conversation duration. The bar graph illustrated in Figure~\ref{fig:compare_word} enables a comparison between the total count of words and sentences inputted by both the agent and the user.

Notably, users tend to engage more extensively with the red agent—characterized by its self-editing and hesitation behavior, resulting in longer interaction duration (Red: $M=238.73 (s), std=224.22$) and higher averages of sentence (Red: $M=10.27 , std=8.31$) and word entries (Red: $M=62.27 , std=46.46$). Conversely, the green agent, embodying only hesitation behaviors, registers the lowest engagement levels from users, evident in the minimal word (Green: $M=49.10, std=27.95$) and sentence (Green: $M=7.0 , std=4.03$) entries and shorter interaction duration (Green: $M=161.60 (s), std=102.31$).

\begin{figure}[t]
    \centering
    \includegraphics[width=0.4\textwidth]{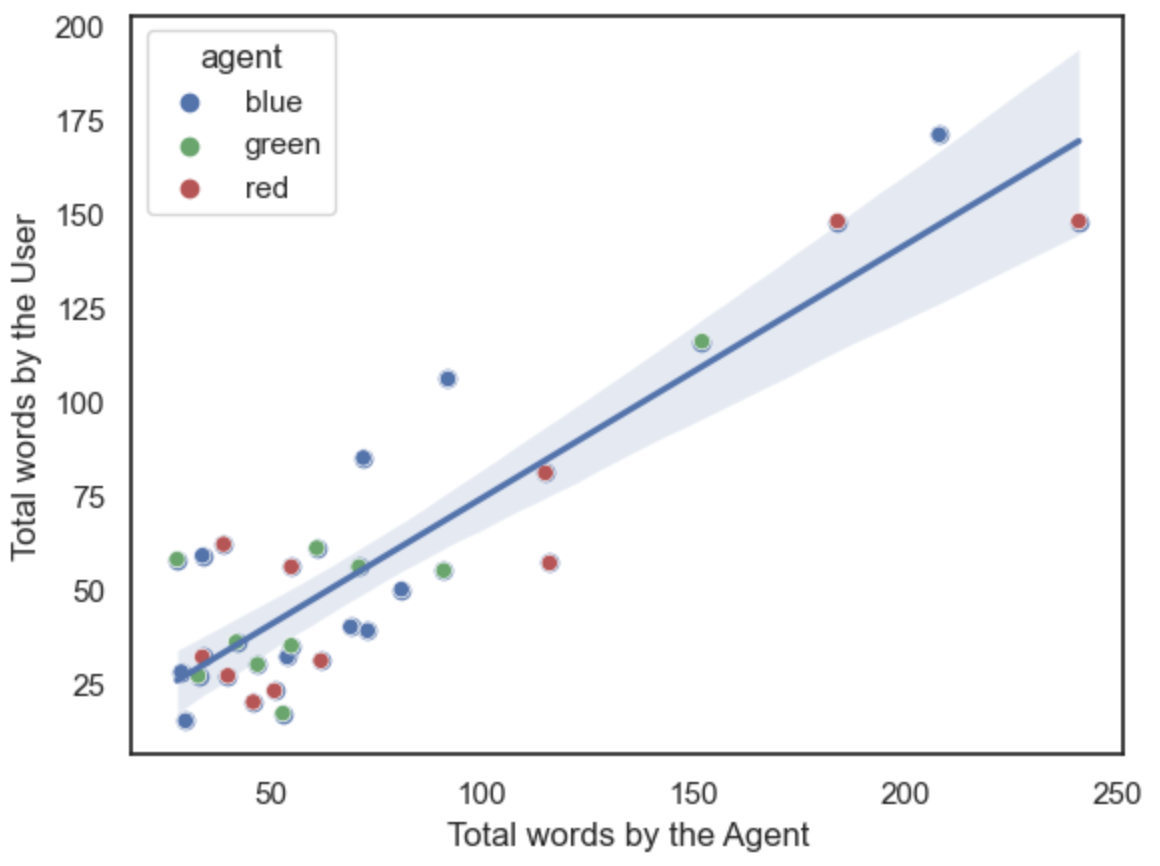}
    \caption{Linear Regression on the count of word entry by the user (Y axis) and by the agent (X axis).}
    \label{fig:regression}
    \Description{The scatter plot shows a linear regression analysis of the count of word entries by the user (Y axis) and by the agent (X axis). Data points are color-coded by agent type: blue, green, and red. The regression line with a shaded confidence interval is displayed, indicating the relationship between the total number of words entered by the user and the agent.}
\end{figure}

Another critical insight pertains to the correlation in the information exchange volume between users and agents. Figure~\ref{fig:regression} demonstrates a positive correlation between the word count entered by the user and that by the agent. The relationship between the total word count by the agent (denoted as $X$) and that by the user (denoted as 
$Y$) is encapsulated in the following linear regression model:

\begin{equation}
    Y = 0.67X + 6.81, R^2 = 0.79
\end{equation}

This equation suggests a substantial linear relationship, with an $R^2$ value of 0.79, indicating that the model explains 79\% of the variance in the user's word count based on the agent’s word count.

\subsection{Qualitative Analysis}
We, two HCI researchers, performed the coding and thematic analysis based on the responses to the open-ended questions in the questionnaire. In the following section, we have consolidated the positive and negative feedback provided by the participants, organizing the comments by agent type. In the following text, participant IDs are noted as per the original recruitment identifiers, ranging from P3 to P22.

\subsubsection{Impression on Blue Agent (baseline)}
Participants expressed a mix of views regarding the Blue agent. On the positive side, participants characterized the agent as “knowledgeable,” “logical,” “enthusiastic,” and “friendly.” One participant (P15) described their interaction with the Blue agent as engaging with a “very responsible and positive” partner. Another participant (P17) appreciated the Blue agent’s communication style, noting that the interaction was “brief, to the point, and light-hearted.”

However, some feedback pointed out areas for improvement. A few participants found the Blue agent to be “too fast” and “not genuine,” with responses that seemed to lack understanding of the context and often appeared to simply mirror the participants’ opinions. For instance, P5 felt the agent’s rapid replies made it seem unreal. Participant P11 and P15 both noted that the agent’s responses could be “too detailed” or “too formal” for casual conversations, making the exchange feel less natural and more rigid. Participant P17 felt the agent was agreeable but in a way that seemed disingenuous, while P4 observed that the level of shared interests between them and the agent seemed “less genuine.” Additionally, participant P8 mentioned that the agent’s lengthy explanation about country music was atypical of human responses in casual chats. Other participants noted inconsistency in language use, a lack of emotional warmth, and a need for the agent to display more character and human-like pausing in its replies to enhance the conversational experience.

\subsubsection{Impression on Green Agent (with only hesitation)}
Participants provided mixed feedback on the Green agent, often highlighting its casual tone while also noting areas for improvement in naturalness and engagement. On one hand, the Green agent was recognized for its casualness, with Participant P11 noting, “I can see him typing, and his replies are more casual than [those of] Blue.”

However, the agent received criticism for not actively extending conversations or understanding context, as seen in comments like, “They did not ask follow-up questions or continue the topic” (P12), and “It cannot talk in context. Questions following the first question it cannot perceive as related topics” (P10). Some participants found the Green agent’s behavior robot-like and unnatural, often repeating information or providing responses incongruent with the discussion. For example, Participant P17 mentioned, “They repeated exactly what I said, which I don't think people do that.” Participant P8 humorously recounted an incongruous response from the agent during a discussion about tennis, observing that the Green agent “don’t act like a human” and “repeats the same answer several times.”

Moreover, several participants pointed out that the Green agent occasionally provided inaccurate responses and failed to understand their questions. Despite these criticisms, some acknowledged that while the Green agent was adept at responding to inquiries, it could be improved as a conversation partner by engaging more with the participant and asking follow-up questions, enhancing the conversational flow and engagement. Participants also suggested that the agent should strive to remember prior knowledge shared in the conversation and respond in a way that facilitates easier replies from the other party. These improvements, they felt, would make interactions with the Green agent feel more like a genuine conversation.

\subsubsection{Impression on Red Agent (with self-editing and hesitation)}
Participants had varying perceptions regarding the Red agent, which exhibited both self-editing and hesitation behaviors. Many participants found it to be “conversational,” “genuine,” and “knowledgeable.” Positive comments highlighted the agent’s human-like traits, with Participant P11 observing, “Red likes to correct his typos, and sometimes even rewrites the previous answer, which feels like he's trying to mimic human behavior.” Despite this, P11 also mentioned that this attempt at mimicking human behavior felt somewhat “creepy” to them. P15 appreciated that Red could “answer sequential questions” and make mistakes, much like a human would. Participant P22 noted the “slower response time” of the Red agent, which gave the impression that it was contemplating its responses, making the interaction “feel more humane” (P20).

However, negative feedback pointed to issues with logic and the use of robotic language. Participant P10 found Red to give “very repetitive answers” and use language “that normal people won't use.” Others noted the conversation felt “one-sided” (P17) and that the agent struggled with “understanding a sentence with logical structure” (P7). Suggestions for future improvement included utilizing “more diverse wording” and providing more detailed responses. Participants also expressed a desire for the agent to ask questions proactively to demonstrate interest in the conversation. For instance, P12 suggested, “It should elaborate on its opinions more rather than just say something is great.” P5 recommended improving the design so that the “intermediate output of the autoregressive model is hidden” as the visible corrections at the end of the sentences seemed unusual to them.

\subsubsection{Comparison between the three agents}
When participants were queried about their preferred agent, responses varied, revealing distinct perceptions and preferences among the three agents. A majority of six participants favored the Red agent, primarily attributing their preference to its human-like qualities. Participants noted that Red felt “more casual”, “genuine”, and appeared “smarter”, fostering a connection that resembled interaction with a human more closely. Meanwhile, three participants opted for the Blue agent, citing its “responsiveness” and “interactivity” as standout features; they also felt that Blue exuded a level of “enthusiasm” that made the conversation engaging. On the other hand, the Green agent was the preferred choice for two participants, who appreciated its “good response time” and “friendly” demeanor, factors that they believed contributed to a positive conversational experience. Each agent, with its unique characteristics, seemed to cater to different expectations and preferences among the participants, underscoring the subjective nature of the user experience in human-agent interaction.
% In this section, complemented by an accompanying supplementary video, we illustrate the typing behaviors after incorporating hesitation and self-editing patterns.

\section{Discussion}

In this section, we delve into design insights derived from our experiment, discuss the challenges encountered during the design process, and touch upon the limitations of our research and potential future work.

\subsection{Design Implications}

In general, the findings indicate that an AI agent’s typing behavior influences human perception. More specifically, the AI agent that integrates both hesitation and self-editing behaviors (represented by the red agent) tends to be slightly more preferred by participants. This variant of the chat agent is perceived as more natural, amiable, and human-like. Participants attributed qualities like 'intelligence' and capability to the red agent, appreciating its self-correction feature which mirrored human behavior. This led participants to enjoy casual interactions with the red agent more compared to the baseline agent (blue), which was perceived as impersonal and artificial due to its rapid responses.
On the other hand, the agent exhibiting only hesitation behavior (depicted by the green agent) was least favored by participants. They perceived it as sluggish and unnatural, engaging with it for the shortest duration compared to the other two agents.
This finding serves as a design recommendation for future researchers, suggesting the incorporation of self-editing behaviors into AI agents as a strategy to positively alter user perceptions. Implementing this feature may foster a communication experience between humans and agents that is not only more favorable but also feels more natural and akin to human-human interaction.

It's also important to highlight that participants' perceptions of the agent were significantly influenced by content-related factors. Many participants observed that the agent seemed incapable of understanding context or recalling prior knowledge, which could be attributed to the experimental setup. In our pursuit of maintaining a manageable volume of chat and preventing ChatGPT from providing overly lengthy responses, we imposed a word limit on each response and reset the model before generating each modification prompt. Consequently, the model might have struggled to remember previous messages or comprehend the context effectively.
Furthermore, participants expressed a desire for a conversational companion that does more than just respond; they wished for an agent that could also initiate dialogue, ask questions, and showcase interest in them. Incorporating this feature could be crucial in future designs, pushing AI agents towards embodying truly social and engaging conversational companions.

\subsection{Design challenges}
\subsubsection{Shift in User Evaluation Criteria}
One key challenge in our quest to identify human-like traits in chatbots is the unpredictability of robot responses. Our reliance on GPT-3.5 as the content source introduced unique challenges.
Users' references to prior topics, expecting contextual comprehension akin to human conversation, occasionally lead to unrelated responses from ChatGPT due to its limited conversational abilities. This disruption affects our results, as participants tend to prioritize response content over the traits we examined: hesitation and self-corrections.

Moreover, our imposed response length constraints occasionally resulted in GPT-3.5 generating lengthy replies even to brief user greetings like 'hi,' contrary to the brevity typically expected in human interactions. Despite our advice to participants to avoid overly concise sentences, customary practices like these can lead to user confusion when the responses appear unnatural. In these situations, users may begin to question the nature of their conversational partner, especially if other unnatural traits become apparent.
Once participants realize that the agent they are chatting with is not a human, they unconsciously shift their evaluation criteria. Their focus shifts towards usability, helpfulness, and heightened expectations for response time and accuracy, potentially contradicting our initial hypotheses.

Another contributing factor to this limitation is our incorporation of self-correction behaviors, wherein we replace the current (incorrect) word with the correct word. While the correct word is sourced from ChatGPT's response, the choice of the incorrect word, whether a redundant term or a synonym, can sometimes appear unnatural. This unnaturalness stems from the challenge of finding an appropriate synonym or redundant word that mimics the accidental mistakes humans commonly make in their day-to-day messaging. Particularly concerning redundant words, we employ a predefined library containing frequently used chat vocabulary. However, the random selection of these words can still result in choices that appear highly irrelevant within the context of the sentence.

\subsubsection{Experiment Scope and Application Scenarios} Our focus was on guiding participants into conversations centered around daily personal subjects, like hobbies. While this approach suits scenarios such as casual chats and general knowledge discussions, it may not fully address contexts where chatbots operate in roles, such as customer service. We opted to gently restrict our conversation topics to everyday scenarios due to the absence of a prompt source capable of swiftly addressing user-specific inquiries, such as questions about credit card referral bonuses. It's important to note that we deliberately refrained from constraining the range or manner of user questions, as ChatGPT's responses might sometimes deviate from typical service interactions, inadvertently revealing its robotic nature.

\subsubsection{Ethical Considerations of Conversational Agents}
Many studies have highlighted ethical concerns with humanizing AI-based conversational agents, including issues related to bias and fairness, misinformation, privacy, data protection, and the potential effects on user empowerment and inclusivity~\cite{wambsganss2021ethical,fossa2022gender}. In our experiment, we meticulously designed procedures to inform participants about the use of the AI agent at the experiment's conclusion. To accurately reflect real-life scenarios, where it's often unclear if one is interacting with a human or AI in contexts like live chats or phone calls, we intentionally kept the agent's nature undisclosed during initial encounters to more authentically simulate real-world interactions. Additionally, we meticulously crafted the interaction tasks and questions to avoid sensitive topics, thereby preventing the generation of any harmful or culturally sensitive content. By steering discussions towards non-sensitive, everyday topics such as hobbies, we sought to mitigate any psychological impacts and aimed to responsibly navigate the complexities of AI interaction, reflecting our overarching commitment to ethical considerations in AI research and application. Moreover, we ensure that all conversation inputs are securely stored on our private server and are not shared with other users, upholding strict privacy standards.

\subsection{Limitations}
While our study sheds light on the interplay between thinking and self-correction in achieving human-like traits, it's important to note some limitations. This paper presents an initial pilot test, and the number of participants recruited was limited. Additionally, our research heavily relies on qualitative analysis, as quantifying human-like traits remains challenging. While our participants provided valuable insights, some responses lacked detail. These limitations suggest opportunities for future studies to explore larger participant pools, employ more quantitative measures, and refine the chatbot's contextual comprehension for more thorough findings.

\subsection{Future Work}

We have identified a discernible trend in the relationship between human-like traits and user preferences when interacting with companion chatbots. To deepen our understanding of how hesitation and self-correction behaviors influence the human-like design of chatbots, future investigations should explore conditions that minimize user bias resulting from the chat's content. This includes expanding research to diverse applications, shedding light on user reactions to these traits in various contexts and through different interaction modalities such as voice and gesture.

Furthermore, for a comprehensive understanding of scenarios where chat agents with human-like traits, such as making and correcting mistakes, are required, an investigation into their effects in diverse applications is warranted. This exploration can shed light on how users react to these traits in various contexts.

While we have incorporated qualitative feedback from participants regarding chatbots with specific human-like traits, there is room for more in-depth exploration of each individual trait to better comprehend their necessity and applicability. Additionally, investigating the impact of allowing some errors to remain uncorrected, as part of a mixed model, can provide insights into the level of human-likeness that resonates with users.

Most notably, we aspire to collect and analyze user typing behavior data using deep learning models. This endeavor will enable us to investigate whether users prefer interacting with chat agents whose typing styles align with their own.

This research can also be extended to study similar human-like nuances in other modalities, such as voice generation. By examining how variations in tone, pauses, and speech patterns can enhance the realism and accessibility of conversational agents, we aim to develop multimodal interactions that are more intuitive and engaging for users.
\section{Conclusion}

This paper examined the impact of hesitation and self-editing behaviors on user perceptions of chatbot naturalness and likability. While it sheds light on these behaviors, it underscores the complexities of human-computer interaction.
Hesitation behaviors, when introduced alone, did not significantly enhance chatbot naturalness and likability, contrary to our initial hypothesis. However, when combined with self-editing behaviors, a noticeable improvement was observed in naturalness, human-likeness, competence, and friendliness. Although these enhancements did not reach statistical significance, they underscore the intricate balance required in designing chatbot behaviors that are both human-like and efficient, highlighting the potential of self-editing as a valuable trait in fostering more human-like interactions.
The influence of response time on user satisfaction further emphasizes the importance of optimizing multimodal interactions to achieve a balance between human-like traits and responsiveness, a critical aspect in enhancing user experiences with CAs. This study's insights into hesitation and self-editing behaviors contribute to our understanding of how subtle behavioral nuances can affect user engagement and satisfaction in conversational interfaces.
By advancing nuanced conversational behaviors within a multimodal framework, we aim to make conversational agents not just more relatable, but truly engaging companions in a variety of applications. 
In conclusion, this study advances our understanding of hesitation and self-editing behaviors in chatbots, with the self-editing behavior showing particular promise. With continued refinement, chatbots can offer more natural and engaging interactions in both chat and other applications, further enhancing human-computer interactions.


\bibliography{sample-base}

\end{document}